\documentclass[final,3p,times,sort&compress]{elsarticle}

\usepackage{amsmath}
\usepackage{amssymb}
\usepackage{epsfig}

\journal{Physics Letters B}

\begin{document}

\begin{frontmatter}



\title{How to measure the linear polarization of gluons in unpolarized proton\\ 
using the heavy-quark pair leptoproduction}


\author[dubna]{A.V.~Efremov}
\ead{efremov@theor.jinr.ru}
\author[yerphi]{N.Ya.~Ivanov\corref{cor1}}
\cortext[cor1]{Corresponding author.}
\ead{nikiv@yerphi.am}
\author[dubna,dubna2]{O.V.~Teryaev}
\ead{teryaev@theor.jinr.ru}

\address[dubna]{Bogoliubov Laboratory of Theoretical Physics, JINR, 141980 Dubna, Russia}
\address[yerphi]{Yerevan Physics Institute, Alikhanian Br.~2, 0036 Yerevan, Armenia}
\address[dubna2]{Veksler and Baldin Laboratory of High Energy Physics, JINR, 141980 Dubna, Russia}

\begin{abstract}
We study the azimuthal $\cos \varphi$ and $\cos 2\varphi$ asymmetries in heavy-quark pair leptoproduction, $lN\rightarrow l^{\prime}Q\bar{Q}X$, as probes of linearly polarized gluons inside unpolarized proton, where the azimuth $\varphi$ is the angle between the lepton scattering plane $(l,l^{\prime})$ and the heavy quark production plane $(N,Q)$. First, we determine the maximal values for the $\cos \varphi$ and $\cos 2\varphi$ asymmetries allowed by  the photon-gluon fusion with unpolarized gluons; these predictions are large, $(\!\!\sqrt{\,3}-1)/2$ and $1/3$, respectively. Then we calculate the contribution of the transverse-momentum dependent gluonic counterpart of the Boer-Mulders function, $h_{1}^{\perp g}$, describing the linear polarization of gluons inside unpolarized proton. Our analysis shows that the maximum values of the azimuthal distributions depend strongly on the gluon polarization; they vary from 0 to 1 depending on $h_{1}^{\perp g}$. We conclude that the azimuthal $\cos \varphi$ and $\cos 2\varphi$ asymmetries in heavy-quark pair leptoproduction are predicted to be large and very sensitive to the contribution of linearly polarized gluons. For this reason, future measurements of the azimuthal distributions in charm and bottom production at the proposed EIC and LHeC colliders seem to be very promising for determination of the linear polarization of gluons inside unpolarized proton.

\end{abstract}


\end{frontmatter}


\section{Introduction and notation} 
\label{1.0}
Search for the polarized quarks and gluons in unpolarized hadrons is of special interest in studies of the spin-orbit couplings of partons and understanding of the proton spin decomposition. The corresponding transverse momentum dependent (TMD) distributions of the transversely polarized quarks, $h_{1}^{\perp q}(\zeta,\vec{k}_{T}^2)$, and linearly polarized gluons, $h_{1}^{\perp g}(\zeta,\vec{k}_{T}^2)$, in an unpolarized nucleon have been introduced in Refs.~\cite{Boer-Mulders} and \cite{Mulders_2001}. Contrary to its quark version (i.e. so-called Boer-Mulders function $h_{1}^{\perp q}$) the TMD density $h_{1}^{\perp g}$ is $T$- and chiral-even. For this reason, like the unpolarized TMD gluon density $f_{1}^{g}\big(\zeta,\vec{k}_{T}^2\big)$, the function $h_{1}^{\perp g}$ can directly be probed in certain  electroproduction experiments.

Azimuthal correlations in heavy quark pair (and dijet) production in unpolarized electron-proton collisions as probes of the density $h_{1}^{\perp g}$ have been considered in Refs.~\cite{Boer_HQ_1,Boer_HQ_2,Boer_HQ_3}.\footnote{Concerning the opportunities to measure the function $h_{1}^{\perp g}$ in unpolarized hadron-hadron collisions, see review \cite{Boer_2015}.} In leading order (LO) in QCD, the complete angular structure of the pair  production cross section has been obtained in terms of seven azimuthal modulations. However, only two of these modulations are really independent; they can be chosen as the $\cos \varphi$ and $\cos 2\varphi$ distributions, where $\varphi$ is the heavy quark (or anti-quark) azimuthal angle \cite{we9}.

In the present paper, we provide the QCD predictions for the $\cos \varphi$ and $\cos 2\varphi$ asymmetries in the  heavy-quark pair production. Our analysis shows that these azimuthal asymmetries (AAs) are expected to be large in wide kinematic ranges and very sensitive to the function $h_{1}^{\perp g}(\zeta,\vec{k}_{T}^2)$. We conclude that the $\cos \varphi$ and $\cos 2\varphi$ distributions could be good probes of the linearly polarized gluons inside unpolarized proton. 

In Refs.\cite{Boer_HQ_1,Boer_HQ_2,Boer_HQ_3}, it was proposed to study the linearly polarized gluons in unpolarized nucleon using the heavy-quark pair production in the reaction
\begin{equation} \label{1}
l(\ell )+N(P)\rightarrow l^{\prime}(\ell -q)+Q(p_{Q})+\bar{Q}(p_{\bar{Q}})+X(p_{X}). 
\end{equation}
To describe this process, the following hadron-level variables are used:
\begin{align}
\bar{S}&=2\left( \ell\cdot P\right), & y&=\frac{q\cdot P}{\ell\cdot P },& T_{1}&=\left(P-p_{Q}\right)^{2}-m^{2},& S&=\left(q+P\right)^{2}  \notag \\
Q^{2}&=-q^{2}, & x&=\frac{Q^{2}}{2q\cdot P},& U_{1}&=\left( q-p_{Q}\right)^{2}-m^{2}, & z&=-\frac{T_{1}}{2q\cdot P},  \label{2}
\end{align}
where $m$ is the heavy-quark mass.

To probe a TMD distribution, the momenta of both heavy quark and anti-quark, $\vec{p}_{Q}$ and $\vec{p}_{\bar{Q}}$, in the process (\ref{1}) should be measured (reconstructed). For further analysis, the sum and difference of the transverse heavy quark momenta are introduced,
\begin{align} \label{3}
\vec{K}_{\perp}&=\frac{1}{2}\left(\vec{p}_{Q\perp}-\vec{p}_{\bar{Q}\perp}\right), &
\vec{q}_{T}&=\vec{p}_{Q\perp}+\vec{p}_{\bar{Q}\perp},
\end{align}
in the plane orthogonal to the direction of the target and the exchanged photon. The azimuthal angles of $\vec{K}_{\perp}$ and $\vec{q}_{T}$ (relative to the the lepton scattering plane projection, $\phi_l=\phi_{l^{\prime}}=0$) are denoted by $\phi_{\perp}$ and $\phi_{T}$,  respectively. 

\begin{figure*}
\begin{center}
\mbox{\epsfig{file=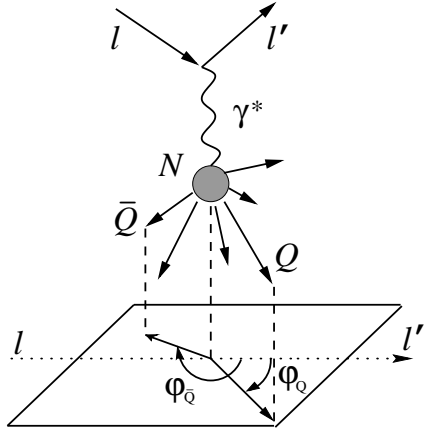,width=200pt}}
\caption{\label{Fg.1}\small Definition of the azimuthal angles $\varphi_Q$ and $\varphi_{\bar{Q}}$ in the nucleon rest frame.}
\end{center}
\end{figure*}

Following Refs.~\cite{Boer_HQ_1,Boer_HQ_2,Boer_HQ_3}, we use the approximation when $\vec{q}_{T}^2\ll \vec{K}_{\perp}^2$ and the outgoing heavy quark and anti-quark are almost back-to-back in the transverse plane, see Fig.~\ref{Fg.1}. In this case, the magnitudes of transverse momenta of the heavy quark and anti-quark are practically the same, $\vec{p}^2_{Q\perp}\simeq \vec{p}^2_{\bar{Q}\perp}\simeq\vec{K}_{\perp}^2$\!. 

At LO, ${\cal O}(\alpha _{em}\alpha_{s})$, the only parton-level subprocess for the reaction (\ref{1}) is the photon-gluon fusion:
\begin{equation} \label{5}
\gamma^{*}(q)+g(k_g)\rightarrow Q(p_{Q})+\bar{Q}(p_{\bar{Q}}), 
\end{equation}
where 
\begin{align} \label{6}
k_{g}^\mu &\simeq \zeta P^\mu + k_{T}^\mu,& \zeta=-\frac{U_1}{y\bar{S}+T_1} &=\frac{q\cdot k_g}{q\cdot P}.
\end{align}
The corresponding parton-level invariants are:
\begin{align} \label{7}
\hat{s}&=(q+k_g)^{2}\simeq\frac{m^2+\vec{K}_{\perp}^2}{z\,(1-z)},& t_{1}&=(k_g-p_{Q})^{2}-m^{2}\simeq\zeta T_1\simeq -z\,(\hat{s}+Q^2),&  u_{1}&=U_1\simeq -(1-z\,)(\hat{s}+Q^2).
\end{align}

\section{Production cross section} 
\label{2.0}

Schematically, the contribution of the photon-gluon fusion to the reaction (\ref{1}) has the following factorized form:
\begin{equation} \label{8}
{\rm d}\sigma\propto L(\ell,q)\otimes \Phi_g(\zeta, k_{T})\otimes \left| H_{\gamma^*g\rightarrow Q\bar{Q}X} (q,k_{g},p_{Q},p_{\bar{Q}})\right|^2, 
\end{equation}
where $L^{\alpha\beta}(\ell,q)=-Q^2 g^{\alpha\beta}+2(\ell^{\alpha}\ell^{\prime\beta}+\ell^{\beta}\ell^{\prime\alpha})$ is the leptonic tensor and $H_{\gamma^*g\rightarrow Q\bar{Q}X}(q,k_{g},p_{Q},p_{\bar{Q}})$ is the amplitude for the hard partonic subprocess. The convolutions $\otimes$ stand for the appropriate integration and traces over the color and Dirac indices.  

Information about parton densities in unpolarized nucleon is formally encoded in  corresponding TMD parton correlators. In particular, the gluon correlator is usually parameterized  as \cite{Mulders_2001}
\begin{equation} \label{9}
\Phi_g^{\mu\nu}(\zeta, k_{T})\propto - g_T^{\mu\nu}f_{1}^{g}\big(\zeta,\vec{k}_{T}^2\big)+\left(g_T^{\mu\nu}-2\frac{k_T^\mu k_T^\nu}{k_T^2}\right)\frac{\vec{k}_{T}^2}{2m^2_N}h_{1}^{\perp g}\big(\zeta,\vec{k}_{T}^2\big), 
\end{equation}   
where $m_N$ is the nucleon mass, 
\begin{align} \label{10}
g_T^{\mu\nu}&=g^{\mu\nu}- \frac{P^\mu n^\nu+P^\nu n^\mu}{P\cdot n},& n^\mu &=\frac{q^\mu+xP^\mu}{P\cdot q} .
\end{align}
In Eq.~(\ref{10}), the tensor $-g_T^{\mu\nu}$ is (up to a factor) the density matrix of unpolarized gluons. The TMD distribution $h_{1}^{\perp g}\big(\zeta,\vec{k}_{T}^2\big)$ describes  the contribution of linearly polarized gluons. The degree of their linear polarization is determined by the quantity $r=\frac{\vec{k}_{T}^2 h_{1}^{\perp g}}{2m^2_N f_{1}^{g}}$. In particular, the gluons are completely polarized along the $\vec{k}_{T}$ direction at $r=1$.\footnote{The TMD densities  under consideration have to satisfy the positivity bound \cite{Mulders_2001}: $\frac{\vec{k}_{T}^2}{2m^2_N}\big|h_{1}^{\perp g}(\zeta,\vec{k}_{T}^2)\big|\leq f_{1}(\zeta,\vec{k}_{T}^2)$.} 

The LO predictions for the azimuth dependent cross section of the reaction (\ref{1}) are presented  in Ref.~\cite{Boer_HQ_2} as follows:
\begin{eqnarray} 
\frac{{\rm d}^{7}\sigma}{{\rm d}y\,{\rm d}x\,{\rm d}z\,{\rm d}^2\vec{K}_{\perp}{\rm d}^2\vec{q}_{T}}={\cal N}\Big\{A_0+A_1\cos\phi_{\perp}+A_2\cos 2\phi_{\perp}+\vec{q}_{T}^2\Big[B_0\cos 2(\phi_{\perp}-\phi_T)+B_1\cos (\phi_{\perp}-2\phi_T) \nonumber \\
+B_1^{\prime}\cos (3\phi_{\perp}-2\phi_T)+B_2\cos 2\phi_T+B_2^{\prime}\cos 2(2\phi_{\perp}-\phi_T)\Big]\Big\}, \label{11}
\end{eqnarray}
where ${\cal N}$ is a normalization factor, $\phi_{\perp}$ and $\phi_T$ denote the azimuthal angles of $\vec{K}_{\perp}$ and $\vec{q}_{T}$, respectively. 
The quantities $A_i$ ($i=0,1,2$) are determined by the unpolarized TMD gluon distribution, $A_i\sim f_{1}^{g}$, while $B_i$ ($i=0,1,2$) and $B_{1.2}^{\prime}$ depend on the linearly polarized gluon density, $B_i^{(\prime)}\sim h_{1}^{\perp g}$.

We have recalculated the cross section for the reaction (\ref{1}) and our results for $A_i$, $B_i$ and $B_{1,2}^{(\prime)}$ do coincide with the corresponding ones presented in Ref.~\cite{Boer_HQ_2}.\footnote{The only exception is an evident misprint in Eq.(25) in Ref.\cite{Boer_HQ_2}. Note also a typo in Eq.(19): instead of ${\rm d}y_i=\frac{{\rm d}z_i}{z_1 z_2}$ should be ${\rm d}y_i=\frac{{\rm d}z_i}{z_i}$.} We have also observed that the expression (\ref{11}) can be simplified essentially in the adopted approximation when $\vec{q}_{T}^2\ll \vec{K}_{\perp}^2$ and the outgoing heavy quark and anti-quark are almost back-to-back in the transverse plane \cite{we9}.

To simplify Eq.~(\ref{11}), it is useful to introduce the sum and difference of magnitudes of the heavy quark transverse momenta,
\begin{align} 
K&=\frac{1}{2}\left(\left|\vec{p}_{Q\perp}\right|+\left|\vec{p}_{\bar{Q}\perp}\right|\right), & \vec{K}_{\perp}^2&=\frac{1}{4}\left(\Delta K\right)^2\sin^2\frac{\alpha}{2}+K^2\cos^2\frac{\alpha}{2}, \notag  \\
\Delta K&=\left|\vec{p}_{Q\perp}\right|-\left|\vec{p}_{\bar{Q}\perp}\right|, & \vec{q}_{T}^2&=\left(\Delta K\right)^2\cos^2\frac{\alpha}{2}+4K^2\sin^2\frac{\alpha}{2}, \label{12}
\end{align}
where $\alpha=\pi-(\varphi_{Q}-\varphi_{\bar{Q}})$ and $\varphi_{Q}$ ($\varphi_{\bar{Q}}$) is the azimuth of the detected quark (anti-quark), see Fig.~\ref{Fg.1}. 

Note first that, at LO in $\alpha_{s}$, the quantity $\Delta K$ is determined by the gluon transverse momentum in the target, i.e. $\Delta K\leq \big|\,\vec{k}_{T}\big|\sim \Lambda_{{\rm QCD}}$ because $\vec{q}_{T}=\vec{k}_{T}$. Then remember that sizable values for the AAs are expected at $K\sim \left|\vec{p}_{Q\perp}\right|\gtrsim m$. In this kinematics (i.e. for $\Delta K\big/K\sim \Lambda_{{\rm QCD}}\big/\,m\ll 1$), the following relations between the azimuthal angles hold:
\begin{align}
\phi_{\perp}&\simeq \frac{\varphi_{Q}+\varphi_{\bar{Q}}}{2}+\frac{\pi}{2}=\varphi_{Q}+ \frac{\alpha}{2},& \alpha &=\pi-(\varphi_{Q}-\varphi_{\bar{Q}}), \notag \\
\phi_{T}&\simeq \frac{\varphi_{Q}+\varphi_{\bar{Q}}}{2}=\varphi_{Q}+\frac{\alpha-\pi}{2},& \phi_{T}&\simeq \phi_{\perp}-\frac{\pi}{2}.  \label{13} 
\end{align}
Corrections to the approximate Eqs.(\ref{13}) are of the order of ${\cal O}(\Delta K\big/K)$. Note also that $\Delta K\big/K\ll 1$ implies $\big|\,\vec{q}_{T}\big|\ll \big|\vec{K}_{\perp}\big|$ for $|\,\varphi_{Q}-\varphi_{\bar{Q}}|\sim \pi$ and vice versa.

One can see from Eqs.~(\ref{13}) that the angles $\phi_{\perp}$ and $\phi_{T}$ are related to  each other at $\Delta K\big/K\ll 1$. This fact means that only two  modulations in Eq.~(\ref{11}) are independent. Integrating Eq.~(\ref{11}) over the anti-quark azimuth, $\varphi_{\bar{Q}}$, we obtain the following cross section:\footnote{The superscript $^{(\pi)}$ on the differential cross section, ${\rm d}^{6}\sigma^{(\pi)}$, means that we perform integration over $\varphi_{\bar{Q}}$ only in region of $|\,\varphi_{Q}-\varphi_{\bar{Q}}|\sim \pi$ where $\big|\,\vec{q}_{T}\big|\ll \big|\vec{K}_{\perp}\big|$ at $\Delta K\big/K\ll 1$. The region of $|\,\varphi_{Q}-\varphi_{\bar{Q}}|\sim 0$ is out of the scope of this paper.}
\begin{eqnarray} 
\frac{{\rm d}^{6}\sigma^{(\pi)}}{{\rm d}y\,{\rm d}x\,{\rm d}z\,{\rm d}\vec{K}_{\perp}^2{\rm d}\vec{q}_{T}^2{\rm d}\varphi}=\frac{e_{Q}^{2}\alpha_{em}^2\alpha_{s}}{8\,\bar{S}^2}\frac{f_{1}^{g}(\zeta,\vec{q}_{T}^2)\hat{B}_2}{y^3 x\,\zeta z\,(1-z)}\Bigg\{\left[1+(1-y)^2 \right]\left(1-2r \frac{\hat{B}^h_2}{\hat{B}_2}\right)-y^2\frac{\hat{B}_L}{\hat{B}_2}\left(1-2r \frac{\hat{B}^h_L}{\hat{B}_L}\right) \nonumber \\
+2(1-y)\frac{\hat{B}_A}{\hat{B}_2}\left(1-2r \frac{\hat{B}^h_A}{\hat{B}_A}\right)\cos2\varphi +(2-y)\sqrt{1-y}\frac{\hat{B}_I}{\hat{B}_2}\left(1-2r \frac{\hat{B}^h_I}{\hat{B}_I}\right)\cos\varphi\Bigg\}, \label{14}
\end{eqnarray}
where $e_Q$ is the heavy quark charge,
\begin{align} \label{15}
\zeta =\frac{-U_1}{y\bar{S}+T_1} &=x+ \frac{m^2+\vec{K}_{\perp}^2}{z\,(1-z)y\,\bar{S}},& r\equiv r(\zeta, \vec{q}_{T}^2)&=\frac{\vec{q}_{T}^2}{2m^2_N}\frac{h_{1}^{\perp g}\big(\zeta,\vec{q}_{T}^2\big)}{f_{1}\big(\zeta,\vec{q}_{T}^2\big)}.
\end{align} 
In Eq.~(\ref{14}), $\varphi$ is the heavy quark azimuth, $\varphi=\varphi_Q$, with the following invariant definition:
\begin{align}
\cos \varphi &=\frac{v\cdot w}{\sqrt{-v^{2}}\sqrt{-w^{2}}},&
\sin \varphi &=\frac{Q^{2}\sqrt{1/x^{2}+4m_{N}^{2}/Q^{2}}}{2\sqrt{-v^{2}}\sqrt{-w^{2}}}~w\cdot \ell, \notag \\
v^{\mu } &=\varepsilon ^{\mu \nu \alpha \beta }P_{\nu }q_{\alpha }\ell _{\beta },& w^{\mu }&=\varepsilon ^{\mu \nu \alpha \beta }q_{\nu }P_{\alpha }p_{Q\beta }.\label{16}
\end{align}

The coefficients $\hat{B}_i$ $(i=2,L,A,I)$ originate from the contribution of unpolarized gluons, while the quantities $\hat{B}^h_i$ are associated with the function $h_{1}^{\perp g}$.
The LO predictions for $\hat{B}_i$ and $\hat{B}^h_i$ $(i=2,L,A,I)$ are given by
\begin{align}
\hat{B}_2\left(z,\vec{K}_{\perp}^2,Q^2\right)&=\frac{1-2l_z}{l_z}+\frac{4\,\hat{k}^2(3l_z+\lambda-1/2)}{(\hat{k}^2+l_z+\lambda )^2},&  \hat{B}^h_2\left(z,\vec{K}_{\perp}^2,Q^2 \right)&=\frac{1}{2}\hat{B}_2\left(z,\vec{K}_{\perp}^2,Q^2\right)-\frac{1-2l_z}{2l_z}, \nonumber \\
\hat{B}_L\left(z,\vec{K}_{\perp}^2,Q^2 \right)&=\frac{8\,\hat{k}^2 l_z}{(\hat{k}^2+l_z+\lambda)^2},& \hat{B}^h_L\left(z,\vec{K}_{\perp}^2,Q^2 \right)&=\frac{1}{2}\hat{B}_L\left(z,\vec{K}_{\perp}^2,Q^2 \right),  \label{17} \\
\hat{B}_A\left(z,\vec{K}_{\perp}^2,Q^2 \right)&=\frac{4\,\hat{k}^2 (l_z+\lambda)}{(\hat{k}^2+l_z+\lambda )^2},& \hat{B}^h_A\left(z,\vec{K}_{\perp}^2,Q^2 \right)&=\frac{1}{2}\hat{B}_A\left(z,\vec{K}_{\perp}^2,Q^2 \right)-1, \nonumber \\
\hat{B}_I\left(z,\vec{K}_{\perp}^2,Q^2 \right)&=\frac{4\,\hat{k}\,(2z-1)}{(\hat{k}^2+l_z+\lambda )^2}(\hat{k}^2-l_z-\lambda ),& \hat{B}^h_I\left(z,\vec{K}_{\perp}^2,Q^2 \right)&=\frac{1}{2}\hat{B}_I\left(z,\vec{K}_{\perp}^2,Q^2 \right), \nonumber
\end{align}
where the following notations are used:
\begin{align} \label{18}
l_z&=z\,(1-z),& \hat{k}^2&=\frac{\vec{K}_{\perp}^2}{Q^2},& \lambda &=\frac{m^2}{Q^2}.
\end{align} 

The quantities $A_i$, $B_i$ ($i=0,1,2$) in Eq.~(\ref{11}) are related to $\hat{B}_k$, $\hat{B}_k^{h}$ ($k=2,L,I,A$) defined by Eqs.~(\ref{17}) as follows:
\begin{align}
A_0&=\kappa_f \left\{[1+(1-y)^2]\hat{B}_2-y^2\hat{B}_L \right\},& A_1&=\kappa_f (2-y)\sqrt{ 1-y}\hat{B}_I, & A_2&=2\kappa_f (1-y)\hat{B}_A, \label{17a} \\
B_0&=\kappa_h\left\{[1+(1-y)^2]\hat{B}_2^h-y^2\hat{B}_L^h \right\},&B_1+B_1^{\prime}&=\kappa_h(2-y)\sqrt{1-y}\hat{B}_I^h,&B_2+B_2^{\prime}&=2\kappa_h (1-y)\hat{B}_A^h, \nonumber
\end{align}
with $\kappa_f=\frac{m^2+\vec{K}_{\perp}^2}{2\xi y\,\bar{S}}e^2_{\!Q}f_1^{g}$ and $\kappa_h=\frac{m^2+\vec{K}_{\perp}^2}{2\xi y\,\bar{S}m_N^2}e^2_{\!Q}\,h_1^{\perp g}$.

\section{Azimuthal $\cos2\varphi$ asymmetry} 
\label{3.0}

One can see from Eq.(\ref{14}) that the gluonic version of the Boer-Mulders function, $h_{1}^{\perp g}$, can, in principle, be determined from measurements of the $\vec{q}_{T}^2$-dependence of the $\cos\varphi$ and $\cos 2\varphi$ asymmetries. Let us first discuss the pQCD predictions for the $\cos 2\varphi$ distribution in the case when the unpolarized gluons only contribute to the reaction (\ref{1}), i.e. for $r=\frac{\vec{q}_{T}^2\, h_{1}^{\perp g}}{2 m^2_N\, f_{1}}=0$. At sufficiently small $y\ll 1$ and fixed values of $Q^2$, the corresponding asymmetry is:
\begin{equation} \label{19}
A_{\cos2\varphi}\left(z,\vec{K}_{\perp}^2\right)\simeq \frac{\hat{B}_A}{\hat{B}_2}\left(z,\vec{K}_{\perp}^2\right)=\frac{4\,\hat{k}^2 l_z\, (l_z+\lambda)}{\hat{k}^4 (1-2l_z)+2\hat{k}^2 (4 l^2_z+\lambda)+(1-2l_z)(l_z+\lambda)^2},
\end{equation} 
where the variables $l_z$, $\hat{k}^2$ and $\lambda$ are defined by Eq.~(\ref{18}).

Our analysis shows that the function $A_{\cos2\varphi}(z,\vec{K}_{\perp}^2)$ has an extremum  at $l_z=1/4$ and $\hat{k}^2=\lambda+1/4$ (i.e. at $z=1/2$ and $\vec{K}_{\perp}^2=m^2+Q^2/4$). This maximum value is:
\begin{equation} \label{20}
A_{\cos2\varphi}\left(z=1/2,\vec{K}_{\perp}^2=m^2+Q^2/4\right)=\frac{1}{3}.
\end{equation}
Eq.~(\ref{20}) implies that the maximum value of the azimuthal $\cos2\varphi$ asymmetry is independent of $m^2$ and $Q^2$, thus it is the same for both charm and bottom quarks at arbitrary values of $Q^2$.
\begin{figure}
\begin{center}
\begin{tabular}{cc}
\mbox{\epsfig{file=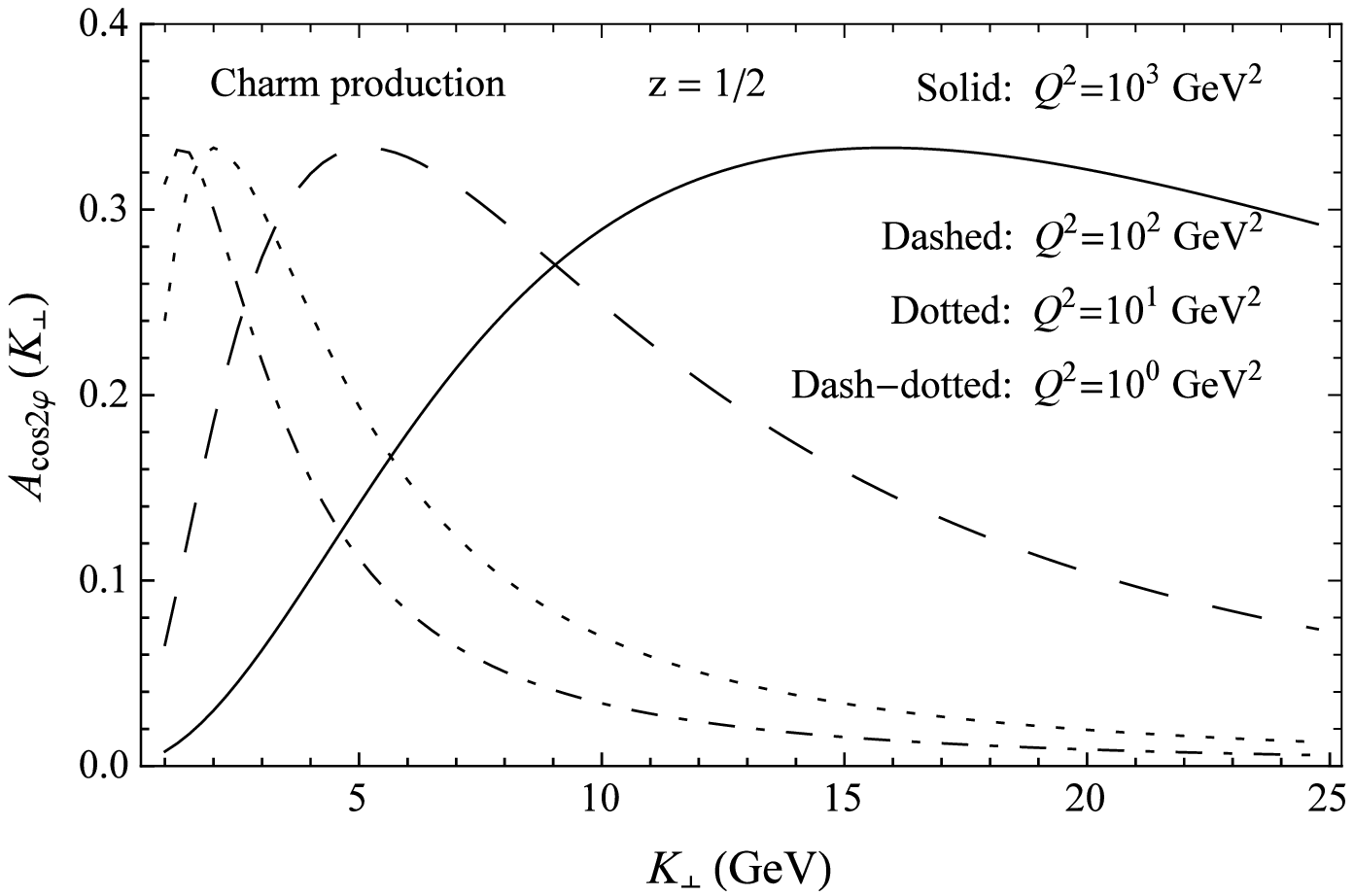,width=220pt}}
& \mbox{\epsfig{file=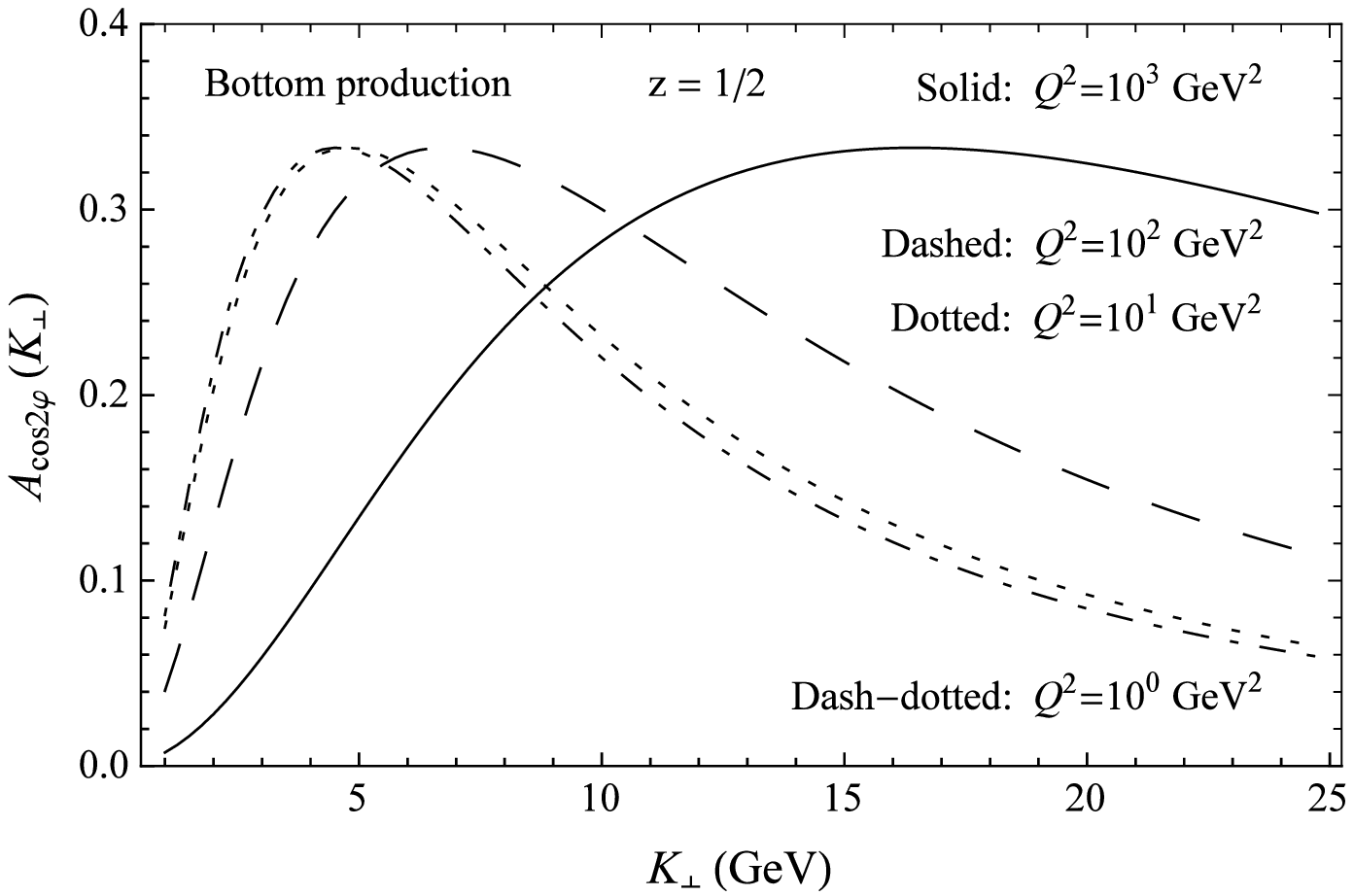,width=220pt}}\\
\end{tabular}
\caption{\label{Fg.2}\small Azimuthal $\cos2\varphi$ asymmetry $A_{\cos2\varphi}(K_{\perp})\equiv A_{\cos2\varphi}(z=1/2,K_{\perp})$ in charm ({\it left panel}) and bottom ({\it right panel}) production as a function of $K_{\perp}=\big|\vec{K}_{\perp}\big|$ at several values of $Q^2$.}
\end{center}
\end{figure}

The LO predictions for the asymmetry $A_{\cos2\varphi}(K_{\perp})\equiv A_{\cos2\varphi}(z=1/2,K_{\perp})$ in charm and bottom production as a function of $K_{\perp}=\big|\vec{K}_{\perp}\big|$ at several values of $Q^2$ are presented in Fig.~\ref{Fg.2}. In the present  paper, we use $m_c=$ 1.25 GeV and $m_b=$ 4.5 GeV. 
One can see from Fig.~\ref{Fg.2} that sizable values for the $\cos2\varphi$ asymmetry are expected in wide regions of $K_{\perp}$ and $Q^2$.

Let us now discuss the contribution of the linearly polarized gluons to the $\cos 2\varphi$ distribution. One can see from Eq.~(\ref{14}) that, at $y\ll 1$ and fixed values of $Q^2$, the corresponding asymmetry $A^h_{\cos2\varphi}(z,\vec{K}_{\perp}^2,r)$, containing the contributions of both $f_1^g$ and $h_1^{\perp g}$ densities, is a function of three variables: 
\begin{equation} \label{21}
A^h_{\cos2\varphi}\left(z,\vec{K}_{\perp}^2,r\right)\simeq \frac{\hat{B}_A}{\hat{B}_2}\frac{1-2r\hat{B}^h_A\Big/\hat{B}_A}{1-2r\hat{B}^h_2\Big/\hat{B}_2}.
\end{equation}
Our analysis shows that the function $A^h_{\cos2\varphi}(z,\vec{K}_{\perp}^2,r)$ has a maximum at $z=1/2$ and $\vec{K}_{\perp}^2=m^2+Q^2/4$ for all values of $r$ in the interval $-1\leq r\leq 1$. We find 
\begin{equation} \label{22}
A^h_{\cos2\varphi}(r)\equiv A^h_{\cos2\varphi}\left(z=1/2,\vec{K}_{\perp}^2=m^2+Q^2/4,r\right)=\frac{1+r}{3-r},
\end{equation}
where $r=\frac{\vec{q}_{T}^2}{2m^2_N}\frac{h_{1}^{\perp g}(\zeta_2,\vec{q}_{T}^2)}{f_{1}(\zeta_2,\vec{q}_{T}^2)}$ describes the degree of the linear polarization of gluons, and $\zeta_2=2x\,(1+4\lambda)$.

\begin{figure}
\begin{center}
\mbox{\epsfig{file=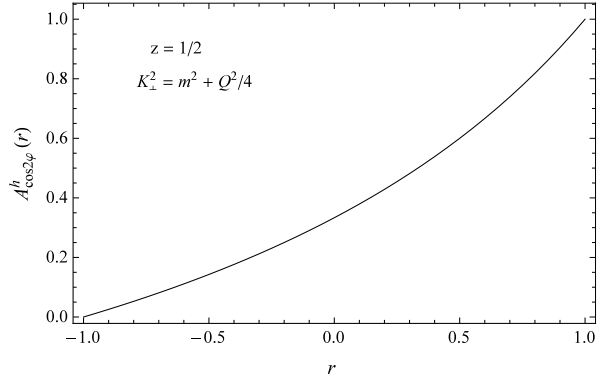,width=220pt}}
\caption{\label{Fg.3}\small Maximum value of the $\cos2\varphi$ asymmetry with the contribution of linearly polarized gluons, $A^h_{\cos2\varphi}(r)$, as a function of $r$, see Eqs.~(\ref{21}), (\ref{22}).}
\end{center}
\end{figure}

One can see from Eq.~(\ref{22}) that the maximum value of the $\cos2\varphi$ asymmetry with the contribution of linearly polarized gluons, $A^h_{\cos2\varphi}(r)$, is a simple function of only variable $r$ (i.e. it is independent of $m^2$ and $Q^2$). The function $A^h_{\cos2\varphi}(r)$ is depicted in Fig.~\ref{Fg.3} where its strong dependence on the variable $r$ is seen. In particular, the upper bound on the $\cos2\varphi$ distribution is equal to 1, while the lower one vanishes. 

In Fig.~\ref{Fg.3}, the $\cos 2\varphi$ asymmetry is depicted at $z=1/2$, $\vec{K}_{\perp}^2=m^2+Q^2/4$ where it predicted to be maximal within pQCD. Note however that the gluon polarization can, in principle, be determined experimentally from measurements of the $\vec{q}_{T}^2$-dependence of the asymmetry in arbitrary kinematic. Our analysis shows that the quantity $A^h_{\cos 2\varphi}(z,\vec{K}_{\perp}^2,r)$ defined by Eq.~(\ref{21}) is an unambiguous function of $r$ for any (physical) values of $z$, $\vec{K}_{\perp}^2$ and $Q^2$. For this reason, comparing future experimental values of the $\cos 2\varphi$ distribution, $A^{\,\rm exp}_{\cos 2\varphi}(z,\vec{K}_{\perp}^2,\vec{q}_{T}^2)$, with the predicted ones, $A^{\,\rm exp}_{\cos 2\varphi}(z,\vec{K}_{\perp}^2,\vec{q}_{T}^2)=A^h_{\cos 2\varphi}(z,\vec{K}_{\perp}^2,r)$, one could unambiguously estimate the quantity $r=\frac{\vec{q}_{T}^2\, h_{1}^{\perp g}}{2 m^2_N\, f_{1}}$ for given $z$, $\vec{K}_{\perp}^2$, $Q^2$ and $\vec{q}_{T}^2$.\footnote{Due to confinement, the heavy quark momenta can be determined/reconstructed from data with an accuracy not better than ${\cal O}\left(\Lambda_{\rm QCD}\right)$. For this reason, too small values of $|\vec{q}_{T}|\sim \Lambda_{\rm QCD}$ seem to be inaccessible.} We conclude that this asymmetry can be good probe of the gluonic counterpart of the Boer-Mulders function, $h_{1}^{\perp g}(\zeta,\vec{q}_{T}^2)$. 

\section{Azimuthal $\cos\varphi$ asymmetry} 
\label{4.0}
At small $y\ll 1$ and fixed values of $Q^2$, the azimuthal $\cos\varphi$ asymmetry due to the  contribution of unpolarized gluons only has the following form:  
\begin{equation} \label{23}
A_{\cos\varphi}\left(z,\vec{K}_{\perp}^2\right)\simeq \frac{\hat{B}_I}{\hat{B}_2}\left(z,\vec{K}_{\perp}^2\right)=\frac{4\,\hat{k}\, (2z-1) l_z\, (\hat{k}^2-l_z-\lambda)}{(1-2l_z)(\hat{k}^2+l_z+\lambda)^2+4\,\hat{k}^2 l_z\,(3 l_z+\lambda-1/2)}. 
\end{equation}  
Contrary to the $\cos2\varphi$ distribution, the quantity $A_{\cos\varphi}(z,\vec{K}_{\perp}^2)$ is an alternating function of both $z$ and $\vec{K}_{\perp}^2$. In particular, $A_{\cos\varphi}(z,\vec{K}_{\perp}^2)=-A_{\cos\varphi}(1-z,\vec{K}_{\perp}^2)$ and $A_{\cos\varphi}(z,\vec{K}_{\perp}^2)=-A_{\cos\varphi}(z,\,(l_z+\lambda)^2Q^4\!\big/\vec{K}_{\perp}^2)$. The function $A_{\cos\varphi}(z,\vec{K}_{\perp}^2)$ vanishes in the region where the $\cos2\varphi$ distribution takes its maximum values: $A_{\cos\varphi}(z=1/2,\vec{K}_{\perp}^2)=A_{\cos\varphi}(z,\vec{K}_{\perp}^2=Q^2z\,(1-z)+m^2)=0$. After integration over the entire region of $z$, the asymmetry also vanishes: $\int{\rm d}z\, A_{\cos\varphi}(z,\vec{K}_{\perp}^2)=0$.

Our analysis shows that the function (\ref{23}) has four extrema in the physical region of $z$ and $\vec{K}_{\perp}^2$: two maxima and two minima. We describe the extrema points in the ($z,\hat{k}^2$) plane with the help of four functions of $\lambda$: $z_{\pm}\equiv z_{\pm}(\lambda)$ and $\hat{k}^2_{\pm}\equiv\hat{k}^2_{\pm}(\lambda)$. The $\cos\varphi$ asymmetry takes its maximum and minimum values at ($z_{\pm},\hat{k}^2_{\pm}$) and ($z_{\pm},\hat{k}^2_{\mp}$), respectively. 

The functions $z_{\pm}(\lambda)$ and $\hat{k}^2_{\pm}(\lambda)$  are given by
\begin{align} 
z_{\pm}&=\frac{1}{2}\left(1\pm\sqrt{1-4l_0}\, \right), \label{24} \\
\hat{k}^2_{\pm}&=\frac{1}{2(1-2l_0)}\left(\!\sqrt{l_0(2l_0+1)+2\lambda(1-l_0)}\pm \sqrt{l_0(3-2l_0)+2\lambda(2-3l_0)}\, \right)^2, \label{25}
\end{align} 
where 
\begin{align} 
l_0&=\frac{2(1-6\lambda)}{3(1-4\lambda)}+\frac{\sqrt{\phantom{\big[}13-18\lambda\,\left(5-12\lambda\right)}}{6(1-4\lambda)}\left(\cos\beta-\sqrt{3}\sin\beta \right), \label{26} \\ 
\beta&=\frac{1}{3}\arctan\left(\frac{9(1-4\lambda)\sqrt{(1+4\lambda)\,\big[1+54\lambda\,[1-6\lambda\,(1-2\lambda)]\big]}}{-46+108\lambda\,\big[5-6\lambda\,(3-4\lambda)\big]} \right). \nonumber
\end{align} 

\begin{figure}
\begin{center}
\begin{tabular}{cc}
\mbox{\epsfig{file=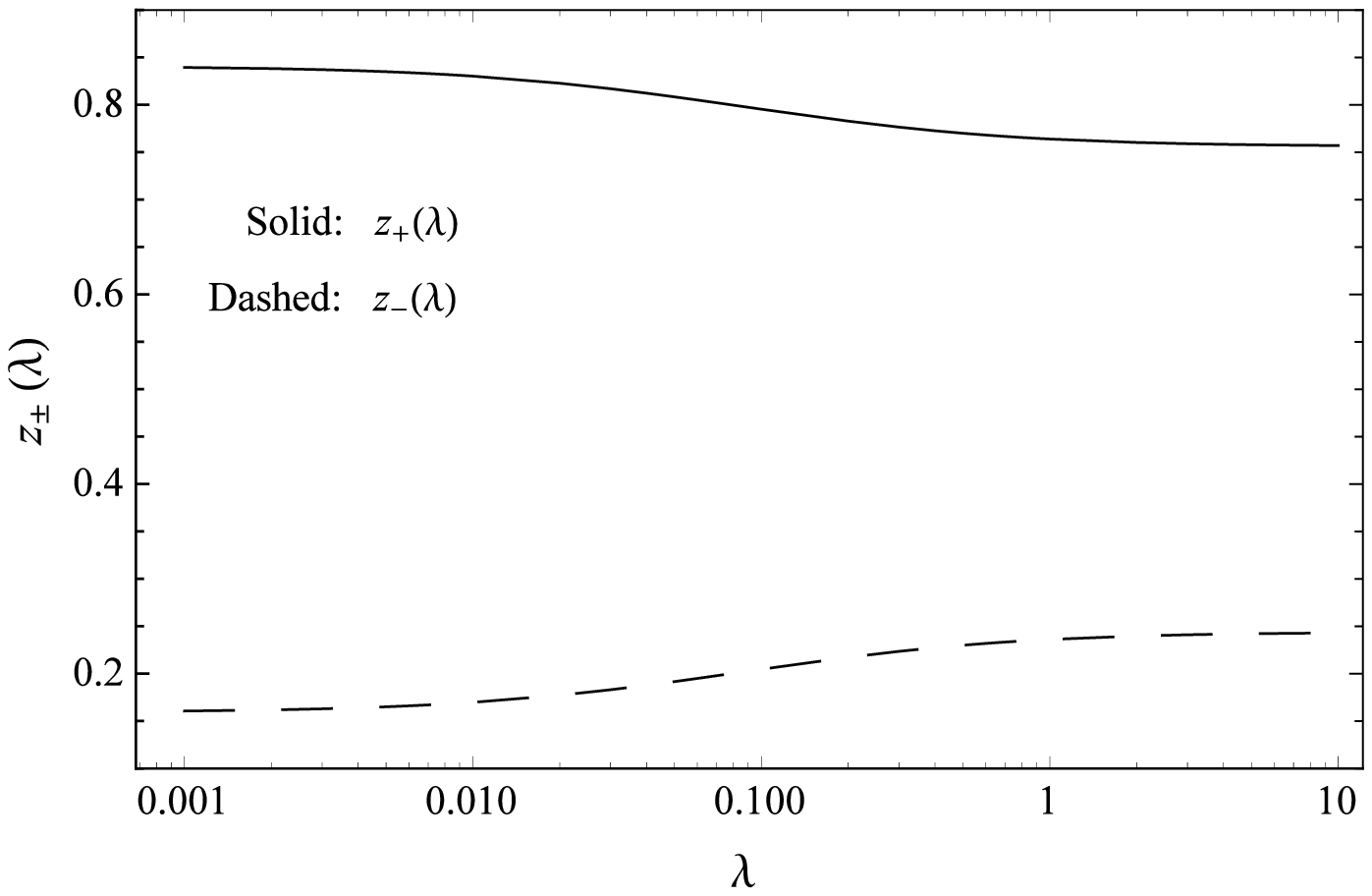,width=220pt}}
& \mbox{\epsfig{file=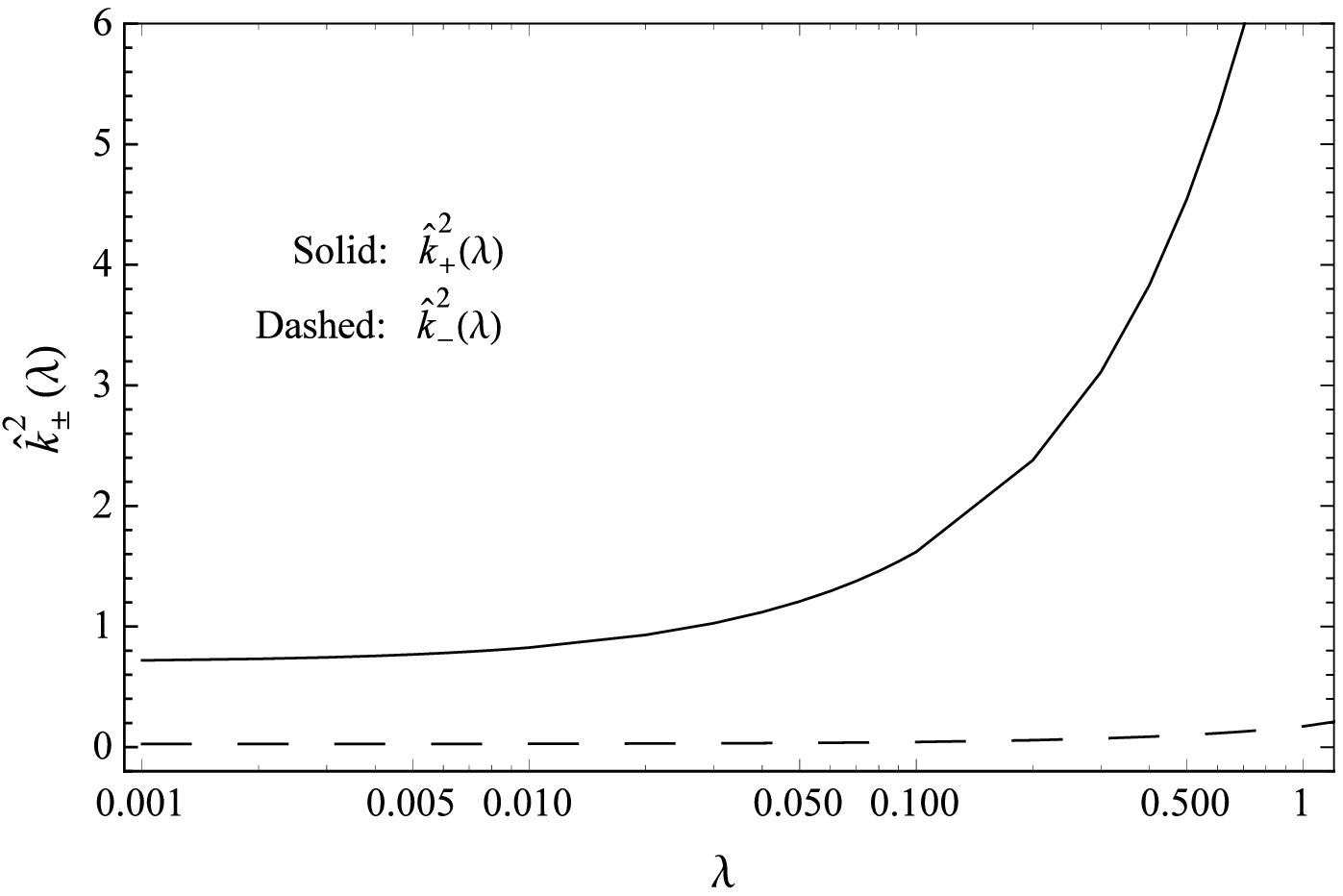,width=220pt}}\\
\end{tabular}
\caption{\label{Fg.4}\small Extrema points of the $\cos\varphi$ asymmetry, $z_\pm$ ({\it left panel}) and $\hat{k}^2_{\pm}$ ({\it right panel}), as functions of $\lambda$, see Eqs.~(\ref{24}), (\ref{25}).}
\end{center}
\end{figure}

The function $l_0\equiv l_0(\lambda)$ depends weakly on $\lambda$; it varies from $l_0(\lambda\rightarrow 0)=1-\sqrt{3}\big/2\simeq 0.134$ to $l_0(\lambda\rightarrow \infty)\simeq 0.184$.
In Fig.~\ref{Fg.4}, the $\lambda$-dependence of the quantities $z_\pm$ (left panel) and $\hat{k}^2_{\pm}$ (right panel) is shown. 
At high $Q^2\gg m^2$, the extrema points are 
\begin{align}\label{27}
z_\pm(\lambda\rightarrow 0)&=\frac{1}{2}\left[1\pm\sqrt[4]{3\big/4}\left(\!\sqrt{3}-1 \right)\right]\simeq\genfrac{\{}{.}{0pt}{0}{\,0.841}{\,0.159},& \hat{k}^2_{\pm}(\lambda\rightarrow 0)&=\frac{\sqrt{3}-1}{2}\left[1\pm\sqrt[4]{3\big/4}\,\right]\simeq\genfrac{\{}{.}{0pt}{0}{\,0.707}{\,0.025}.
\end{align}

\begin{figure}
\begin{center}
\begin{tabular}{cc}
\mbox{\epsfig{file=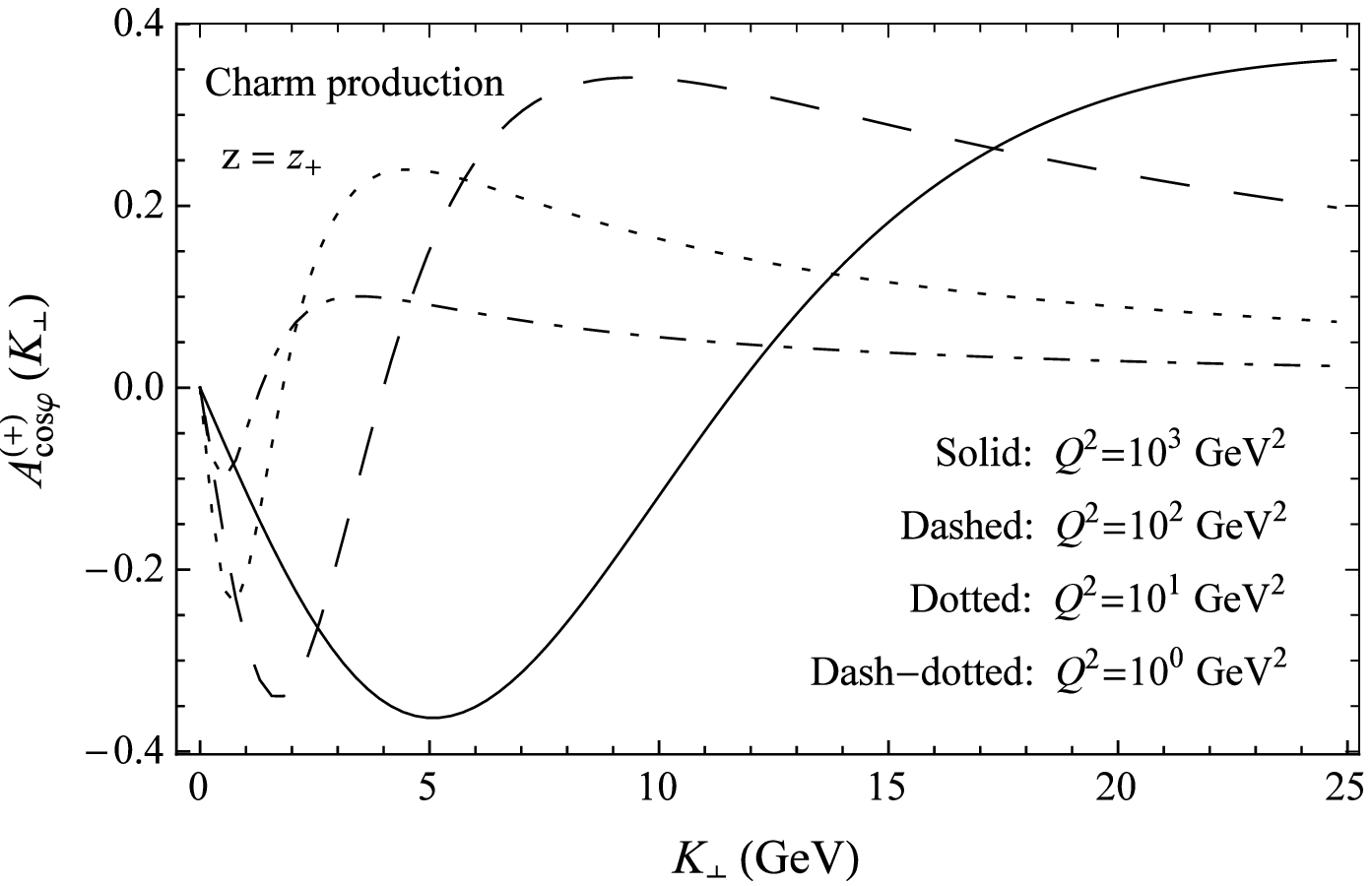,width=220pt}}
& \mbox{\epsfig{file=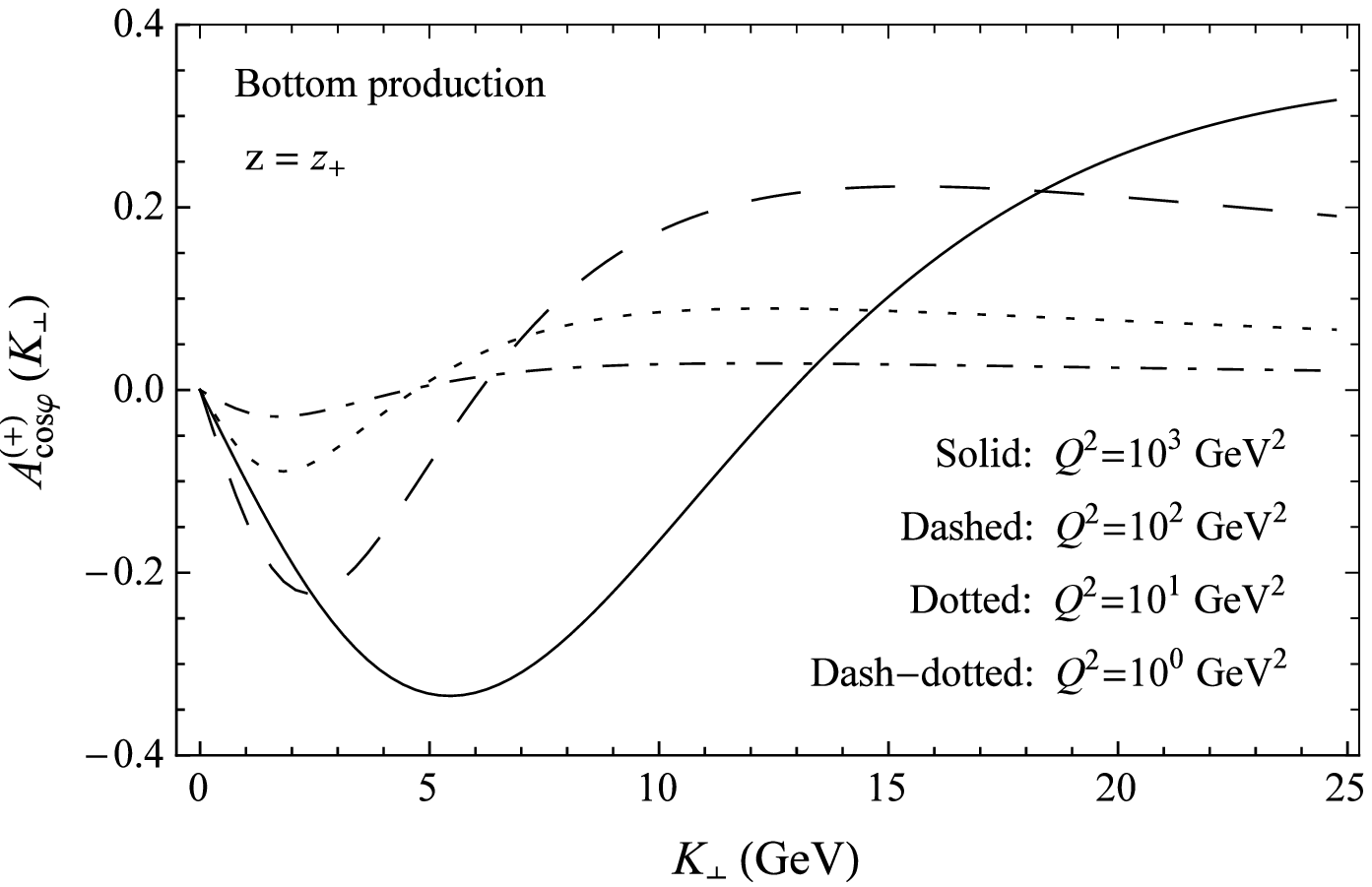,width=220pt}}\\
\end{tabular}
\caption{\label{Fg.5}\small Azimuthal $\cos\varphi$ asymmetry $A^{(+)}_{\cos\varphi}(K_{\perp})\equiv A_{\cos\varphi}(z=z_{+},K_{\perp})$ in charm ({\it left panel}) and bottom ({\it right panel}) production as a function of $K_{\perp}=\big|\vec{K}_{\perp}\big|$ at several values of $Q^2$.}
\end{center}
\end{figure}

The azimuthal $\cos\varphi$ asymmetry $A^{(+)}_{\cos\varphi}(K_{\perp})$ in charm (left panel) and bottom (right panel) production is depicted in Fig.~\ref{Fg.5} as a function of $K_{\perp}=\big|\vec{K}_{\perp}\big|$ at several values of $Q^2$. The quantities $A^{(\pm)}_{\cos\varphi}(K_{\perp})$ are defined as
\begin{align}\label{28}
A^{(\pm)}_{\cos\varphi}(K_{\perp})\equiv A_{\cos\varphi}(z=z_{\pm},K_{\perp}).
\end{align}
Since $z_{+}+z_{-}=1$, $A^{(-)}_{\cos\varphi}(K_{\perp})=-A^{(+)}_{\cos\varphi}(K_{\perp})$.   
One can see from Fig.~\ref{Fg.5} that the maximum (and minimum) value of the asymmetry grows with $Q^2$. 

At arbitrary $m^2$ and $Q^2$, the quantities $A_{\cos\varphi}(z=z_{\pm},\vec{K}_{\perp}^2=\hat{k}^2_{\pm}Q^2)$ are complicated functions of $\lambda$ which are too long to be presented here. As functions of $\lambda$, these quantities take their maximum and minimum values at  $\lambda\rightarrow 0$:
\begin{align}
A_{\cos\varphi}\left(z=z_{+}(\lambda),\vec{K}_{\perp}^2=Q^2\hat{k}^2_{+}(\lambda)\right)&=A_{\cos\varphi}\left(z=z_{-}(\lambda),\vec{K}_{\perp}^2=Q^2\hat{k}^2_{-}(\lambda)\right)\stackrel{\lambda\rightarrow 0}{=}\frac{\sqrt{3}-1}{2}\simeq 0.366, \label{29} \\
A_{\cos\varphi}\left(z=z_{+}(\lambda),\vec{K}_{\perp}^2=Q^2\hat{k}^2_{-}(\lambda)\right)&=A_{\cos\varphi}\left(z=z_{-}(\lambda),\vec{K}_{\perp}^2=Q^2\hat{k}^2_{+}(\lambda)\right)\stackrel{\lambda\rightarrow 0}{=}-\frac{\sqrt{3}-1}{2}\simeq -0.366. \nonumber
\end{align}
So, this value of the $\cos\varphi$ asymmetry, $(\!\sqrt{3}-1)/2$, is the maximal one allowed by the photon-gluon fusion with unpolarized initial gluons. 

Let us now consider the contribution of the linearly polarized gluons inside unpolarized nucleon to the $\cos\varphi$ distribution. According to Eq.~(\ref{14}), the corresponding asymmetry  $A^h_{\cos\varphi}(z,\vec{K}_{\perp}^2,r)$, containing the contributions of both $f_1^g$ and $h_1^{\perp g}$ densities, at $y\ll 1$ and fixed values of $Q^2$ is described by
\begin{equation} \label{30}
A^h_{\cos\varphi}\left(z,\vec{K}_{\perp}^2,r\right)\simeq \frac{\hat{B}_I}{\hat{B}_2}\frac{1-2r\hat{B}^h_I\Big/\hat{B}_I}{1-2r\hat{B}^h_2\Big/\hat{B}_2}.
\end{equation}
We denote the values of this function at $(z_\pm,\hat{k}^2_{\pm})$ and $(z_\pm,\hat{k}^2_{\mp})$ as $A^{h(+)}_{\cos\varphi}(r)$ and $A^{h(-)}_{\cos\varphi}(r)$, respectively:\footnote{Note that $A^{h(-)}_{\cos\varphi}(r)=-A^{h(+)}_{\cos\varphi}(r)$ at the same values of $\lambda$. }
\begin{align} \label{31}
A^{h(+)}_{\cos\varphi}(r)&\equiv A^h_{\cos\varphi}\left(z=z_{\pm},\vec{K}_{\perp}^2=Q^2\hat{k}^2_{\pm},r\right),& A^{h(-)}_{\cos\varphi}(r)&\equiv A^h_{\cos\varphi}\left(z=z_{\pm},\vec{K}_{\perp}^2=Q^2\hat{k}^2_{\mp},r\right).
\end{align}
Contrary to the quantity $A^{h}_{\cos2\varphi}(r)$ given by Eq.~(\ref{22}), the functions $A^{h(\pm)}_{\cos\varphi}(r)$ depend explicitly on $\lambda$. For high $Q^2\gg m^2$, we have
\begin{align} \label{32}
A^{h(\pm)}_{\cos\varphi}(r)\stackrel{\lambda\rightarrow 0}{=}\pm \frac{\big(\!\sqrt{3}-1\big)\,\big(1-r\big)}{2-r\,\big(1-2\big/\sqrt{3}\big)},
\end{align}
where $r=\frac{\vec{q}_{T}^2}{2m^2_N}\frac{h_{1}^{\perp g}}{f_{1}}(\zeta_{1},\vec{q}_{T}^2)$,  while $\zeta_{1}$ is determined by the quantity $\hat{k}^2_{\pm}$: $\zeta_{1}=x\,\big(1+\hat{k}^2_{\pm}\big/\,l_0\big)=x\,\big(2+\sqrt{3}\big)\big(1\pm\sqrt[4]{3}\big)\simeq x\cdot\genfrac{\{}{.}{0pt}{1}{\,6.275}{\,1.090}$.
\begin{figure}
\begin{center}
\begin{tabular}{cc}
\mbox{\epsfig{file=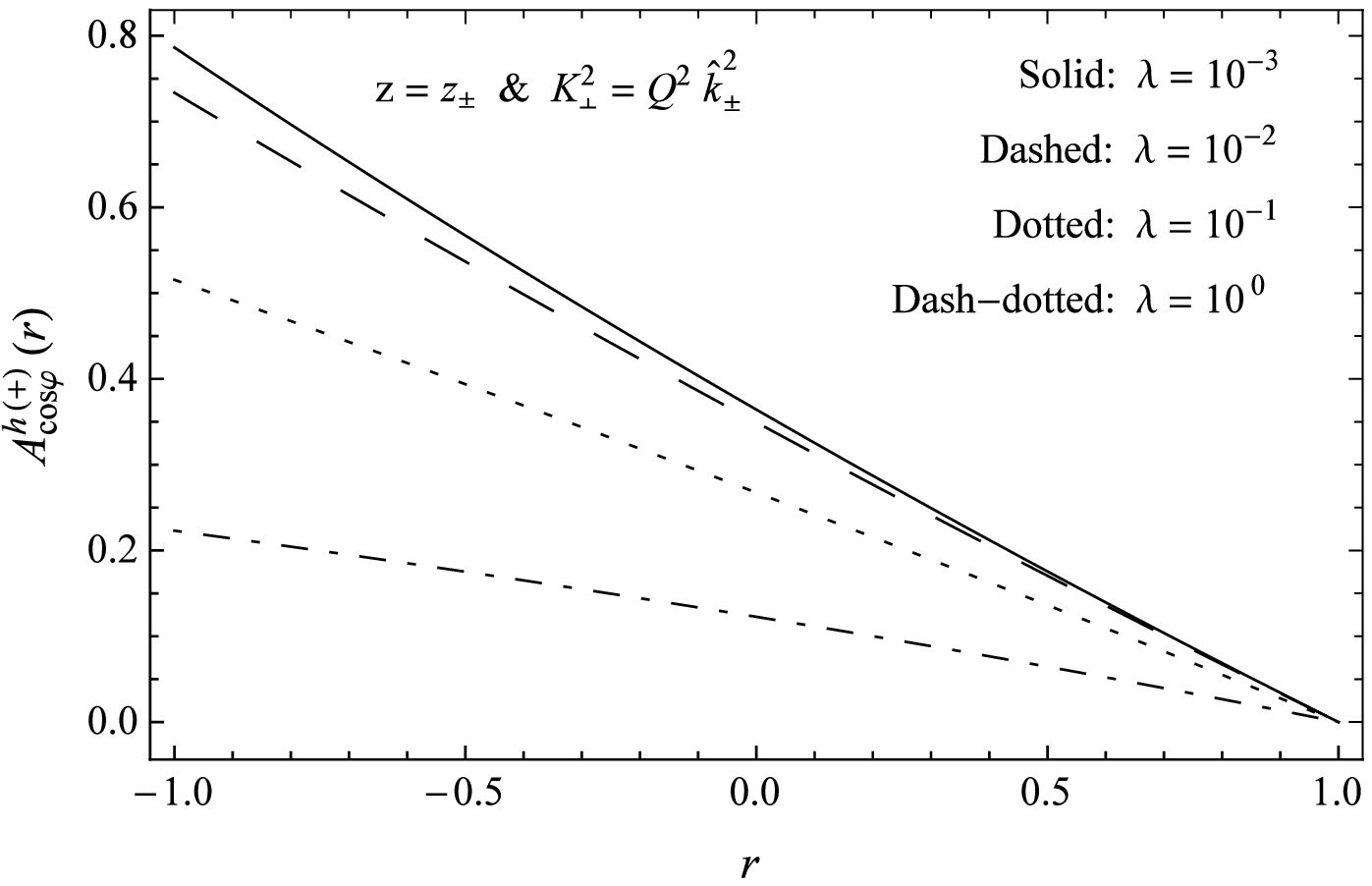,width=220pt}}
& \mbox{\epsfig{file=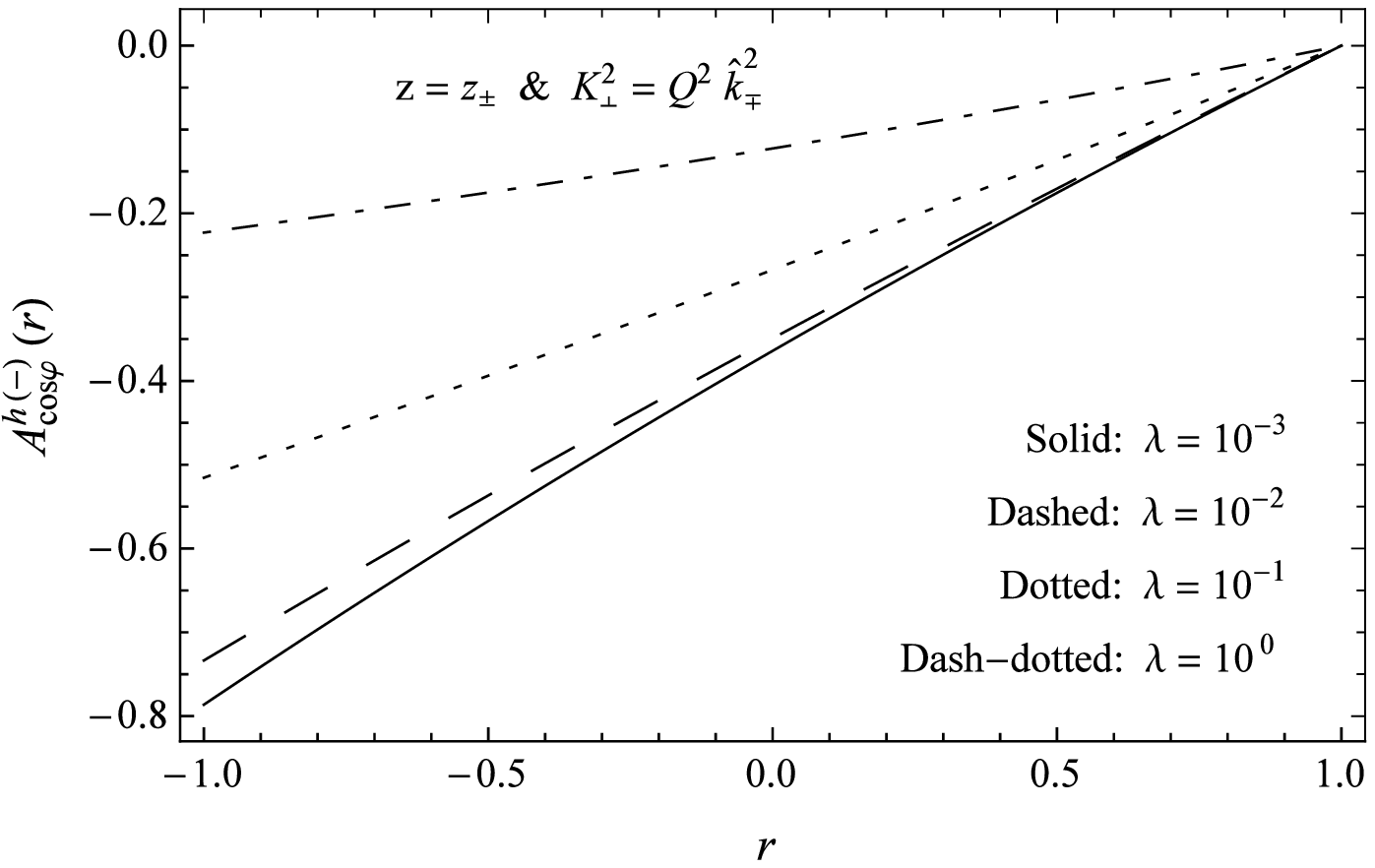,width=220pt}}\\
\end{tabular}
\caption{\label{Fg.6}\small Azimuthal $\cos\varphi$ asymmetry with the contribution of linearly polarized gluons at $(z=z_{\pm},K_{\perp}^2=Q^2\hat{k}^2_{\pm})$ ({\it left panel}) and $(z=z_{\pm},K_{\perp}^2=Q^2\hat{k}^2_{\mp})$ ({\it right panel}) as a function of $r$ at several values of $\lambda$, see Eqs.~(\ref{30}), (\ref{31}).}
\end{center}
\end{figure} 

In Fig.~\ref{Fg.6}, the $\cos\varphi$ asymmetry with the contribution of linearly polarized gluons at $(z=z_{\pm},\vec{K}_{\perp}^2=Q^2\hat{k}^2_{\pm})$ (left panel) and $(z=z_{\pm},\vec{K}_{\perp}^2=Q^2\hat{k}^2_{\mp})$ (right panel) is depicted as a function of $r$ at several values of $\lambda$. One can see from Fig.~\ref{Fg.6}  strong dependence of the asymmetry on both $r$ and $\lambda$. In particular, the upper bound on the absolute value of the functions $A^{h(\pm)}_{\cos\varphi}(r)$ is equal to $\frac{2(\!\sqrt{3}-1)}{3-2/\sqrt{3}}\simeq  0.793$ (the case of $r\rightarrow-1$ and $\lambda\rightarrow 0$), while the lower bound vanishes (the case of $r\rightarrow 1$). 

In Fig.~\ref{Fg.6}, the $\cos \varphi$ asymmetry is depicted at specific values of $z$ and $\vec{K}_{\perp}^2$ where it predicted to be maximal in the case when $h_{1}^{\perp g}=0$. However, similarly to the $\cos 2\varphi$ asymmetry, the quantity $A^h_{\cos \varphi}(z,\vec{K}_{\perp}^2,r)$ defined by Eq.~(\ref{30}) is an unambiguous function of $r$ for any physical kinematics. For this reason, the parameter\, $r=\frac{\vec{q}_{T}^2\, h_{1}^{\perp g}}{2 m^2_N\, f_{1}}$ can, in principle, be unambiguously determined from the equation $A^{\,\rm exp}_{\cos \varphi}(z,\vec{K}_{\perp}^2,\vec{q}_{T}^2)=A^h_{\cos \varphi}(z,\vec{K}_{\perp}^2,r)$ at arbitrary values of the kinematical variables for any $-1< A^{\,\rm exp}_{\cos \varphi}<1$. We conclude that the $\cos\varphi$ asymmetry in heavy-quark pair leptoproduction can also be good probe of the linearly polarized gluons in unpolarized proton.

\section{Conclusion} 
\label{5.0}

In this paper, we study the azimuthal $\cos \varphi$ and $\cos 2\varphi$ asymmetries in heavy-quark pair leptoproduction, $lN\rightarrow l^{\prime}Q\bar{Q}X$, as probes of the linearly polarized gluon density, $h_{1}^{\perp g}(\zeta,\vec{k}_{T}^2)$, inside unpolarized proton. Here the azimuth $\varphi$ is the angle between the lepton scattering plane $(l,l^{\prime})$ and the heavy quark production plane $(N,Q)$.
The maximal values of the $\cos \varphi$ and $\cos 2\varphi$ asymmetries allowed by  the photon-gluon fusion with unpolarized gluons (i.e. when $h_{1}^{\perp g}$=\,0) are $(\!\!\sqrt{\,3}-1)/2$ and $1/3$, respectively. Our analysis shows that azimuthal distributions are very sensitive to the linear polarization of gluons. In particular, the maximum values of both $\cos \varphi$ and $\cos 2\varphi$ asymmetries vary from 0 to 1 depending on $h_{1}^{\perp g}(\zeta,\vec{k}_{T}^2)$.\footnote{The asymmetry $A^h_{\cos\varphi}(z,\vec{K}_{\perp}^2,r)$ defined by  Eq.~(\ref{30}) tends to 1 at $r\rightarrow-1$, $\lambda\rightarrow 0$, $\vec{K}_{\perp}^2\rightarrow 0$ and $z\rightarrow 1$. However, the heavy flavor production rates vanish at the threshold, i.e. for $z\rightarrow 1$. For this reason, we consider the $\cos\varphi$ asymmetry at $z=z_{\pm},\vec{K}_{\perp}^2=Q^2\hat{k}^2_{\pm}$.}

We observe that the contribution of linearly polarized gluons has different impact on the azimuthal distributions. Namely, the positive values of $h_{1}^{\perp g}$ lead to strong  suppression (enhancement) of the $\cos \varphi$ ($\cos 2\varphi$) asymmetry, while the negative ones enhance (suppress) the $\cos \varphi$ ($\cos 2\varphi$) distribution. This fact could have very useful consequence: large experimental values for any of asymmetries, $A^{\,\rm exp}_{\cos2\varphi}>1/3$ or $|A^{\,\rm exp}_{\cos\varphi}|>(\!\!\sqrt{\,3}-1)/2$, will directly indicate not only the presence of linearly polarized gluons but also the sign of $h_{1}^{\perp g}$.

We conclude that both the $\cos \varphi$ and $\cos 2\varphi$ asymmetries in heavy-quark pair  leptoproduction could be good probes of the linear polarization of gluons inside unpolarized nucleon. Concerning the experimental aspects, AAs in charm and bottom production can be measured at the proposed EIC \cite{EIC} and LHeC \cite{LHeC2} colliders.\footnote{As to the current opportunities to measure the azimuthal asymmetries in heavy flavor electroproduction, see Refs.~\cite{we9,we3}.}

Note that the linear polarization of gluons can also be measured using azimuthal correlations in heavy-quark pair and dijet production in unpolarized proton-proton collisions \cite{Boer_dijet,Boer_HQ_1,Boer_HQ_2}. QCD predictions for these correlations in the kinematics of the NICA collider at JINR \cite{NICA} will be considered in a forthcoming publication.

Presently, the AAs in heavy-quark electroproduction are completely unmeasured. Their  (predicted) remarkable properties could however provide very promising applications. In particular, the $\cos 2\varphi$ asymmetry is well defined in pQCD: it is stable both perturbatively and parametrically \cite{we2,we4}. For this reason, it seems to be ideal probe of the heavy-quark densities \cite{we5,we6} (both intrinsic \cite{BHPS,BPS} and perturbative \cite{ACOT,Collins}) and linearly polarized gluon distribution, $h_{1}^{\perp g}$, in unpolarized proton.

\section*{Acknowledgements} The authors are grateful to S.~J.~Brodsky, A.~Kotzinian and   S.~O.~Moch for useful discussions. This work is supported in part by the DAAD foundation, project A/14/04714-326, and State Committee of Science of RA, grant 15T-1C223. 





\end{document}